\documentclass[epj]{svjour}
\usepackage{graphicx}
\usepackage{amssymb}

\newcommand{\dd}{{\rm d}}
\newcommand{\ii}{{\rm i}}

\begin{document}

\title{Multiplicity distributions inside parton cascades developing in a medium}

\author{Nicolas Borghini\inst{1}}
\institute{CERN, Department of Physics, Theory Division, CH-1211 Gen\`eve 23, 
  Switzerland}

\date{Received: date / Revised version: date}
%
\abstract{The explanation of the suppression of high-$p_T$ hadron yields at 
RHIC in terms of jet-quenching implies that the multiplicity distributions of 
particles inside a jet and jet-like particle correlations differ strongly in 
nucleus--nucleus collisions at RHIC or at the LHC from those observed at 
$e^+e^-$ or hadron colliders. 
We present a framework for describing the medium-induced modification, which 
has a direct interpretation in terms of a probabilistic medium-modified parton 
cascade, and which treats leading and subleading partons on an equal footing. 
We show that our approach can account for the strong suppression of single 
inclusive hadron spectra measured in Au--Au collisions at RHIC, and that this 
implies a characteristic distortion of the single inclusive distribution of 
soft partons inside the jet. 
We determine, as a function of the jet energy, to what extent the soft 
fragments within a jet can be measured above some momentum cut. 
} 
\maketitle
\section{Introduction}
\label{intro}

Among the most notable results from the first years of nucleus--nucleus 
collisions at RHIC stand the deficit in high transverse-momentum hadrons and 
the suppression of leading back-to-back hadron correlations observed in central
Au--Au collisions with respect to expectations from scaling the yields measured 
in $pp$ or d--Au collisions~\cite{Jacobs:2004qv}. 
These observations are consistent with the ``jet-quenching'' scenario: 
before they hadronize in the vacuum, partons produced in the dense matter 
created in head-on Au--Au collisions dissipate a significant fraction of their 
energy through an enhanced radiation of soft gluons~\cite{Baier:2000mf,%
  Gyulassy:2003mc,Kovner:2003zj}.

Irrespective of the details of the implementation of the medium-enhanced 
radiation of gluons --- either through coherent multiple soft-momentum transfers
\`a la Landau--Pomeranchuk--Migdal~\cite{Baier:1996sk,Salgado:2003gb}, or 
through single hard scattering~\cite{Gyulassy:2003mc} --- jet-quenching models 
of inelastic (radiative) energy loss are quite successful in explaining present 
light-hadron data from RHIC~\cite{Turbide:2005fk,Dainese:2004te,Eskola:2004cr}.
Nevertheless, there remains much room for technical improvement over the 
existing formulations of inelastic energy loss. 
Thus, a generic feature of these approaches is that they only consider the 
medium-induced enhancement in gluon radiation for the leading parton, 
discarding the medium influence on subleading partons. 
Such an approximation may remain under control when dealing with leading-hadron 
production; however, predictions involving subleading particles become 
questionable, be it for jet shapes, which may become experimentally accessible 
at the LHC~\cite{Mercedes}, or simply for intrajet two-particle correlations.
Similarly, in existing models energy-momentum conservation is not explicitly 
conserved at each parton splitting, but only globally, through various {\em ad 
  hoc\/} corrections. 

A novel formulation of medium-induced parton energy loss was recently 
introduced in Ref.~\cite{Borghini:2005em}, which aims at correcting some of the 
shortcomings of standard approaches mentioned above. 
Thus, it is the first one that deals equally with the various splittings of both
leading and subleading partons inside a shower. 
Furthermore, it automatically conserves energy-momentum at each parton 
splitting. 
These improvements are admittedly obtained, at the moment, at the cost of a less
accurate treatment of other aspects of radiative energy loss, which should be 
restored in a future Monte-Carlo implementation. 
Nevertheless, the physical constraints we choose to emphasize in our approach 
may prove to be of greater importance than those retained in other existing 
models, in particular for the discussion of high-$p_T$ particle correlations 
(see, however, Refs.~\cite{Majumder:2004wh,Majumder:2004pt} for an alternative 
extension of parton energy loss which allows one to tackle that issue) or 
multiplicity distributions inside jets.

\section{Formalism}
\label{sec:formalism}

\begin{figure}[t]
  \includegraphics*[width=\linewidth]{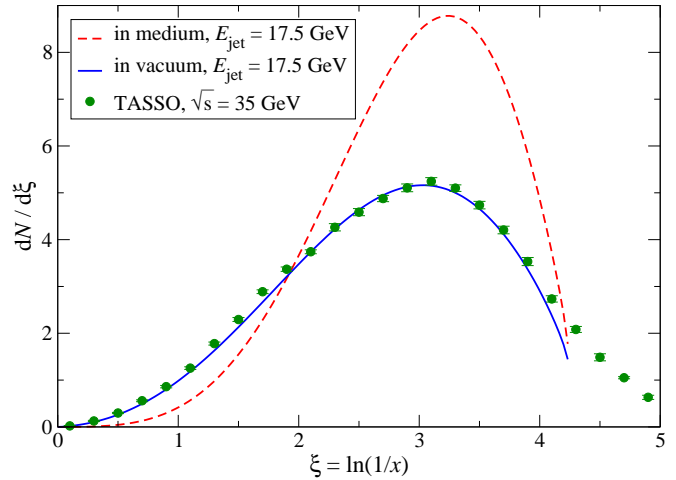}
  \caption{Longitudinal distribution ${\rm d}N/{\rm d}\ln(1/x)$ of inclusive 
    hadrons inside a jet of energy $E_{\rm jet}=17.5$~GeV, as a function of 
    $\ln(1/x)=\ln(E_{\rm jet}/p)$, as measured by TASSO~\cite{Braunschweig:1990yd}
    and within MLLA (solid curve: $f_{\rm med}=0$; dashed curve: 
    $f_{\rm med}=0.8$).}
  \label{fig:hump-backed_plateau}
\end{figure}

One of the most testing ground of the color structure of quantum chromodynamics
(QCD) is provided by the jets that are created in elementary-particle 
collisions, be it in $e^+e^-$ or in $pp$/$p\bar p$ collisions. 
Thus, the asymptotic shape of the distribution of hadron momenta inside a jet 
can be computed exactly, especially at small momentum fractions 
$x=p/E_{\rm jet}$, by a resummation of infrared-singular terms to all orders, 
within the so-called Modified Leading Logarithmic Approximation (MLLA) of 
QCD~\cite{Mueller:1982cq,Bassetto:1984ik,Dokshitzer:1988bq}.
Color coherence thus results in destructive interference between partons, which
leads to a suppression of hadrons with small $x$. 
The interference is actually equivalent, to double and single logarithmic 
accuracy in $\ln(1/x)$ and $\ln(Q/\Lambda_{\rm eff})$ --- where $Q\sim E_{\rm jet}$ 
is the jet virtuality and $\Lambda_{\rm eff}$ an infrared cutoff which is 
eventually fitted to experimental data --- to an angular ordering of the 
sequential parton decays within the shower, with leading-order splitting 
functions. 
An important prediction of this angular-ordered probabilistic parton cascade is,
to next-to-leading order $\sqrt{\alpha_S}$, the characteristic ``hump-backed 
plateau'' shape of the distribution of parton momenta inside a jet, represented 
as a function of $\ln(1/x)$. 
The parton shower, evolved down to an infrared cutoff $Q_0\sim\Lambda_{\rm eff}$, 
is eventually identified to a hadron jet, by mapping locally each parton onto a 
hadron (``Local Parton--Hadron Duality'', LPHD): for each hadron type, the 
hadron distribution equals $K^h$ times the parton distribution, where $K^h$ is 
a proportionality factor of order unity.\footnote{%
  In addition to using different phenomenological proportionality constants 
  $K^h$, the hadron-type dependence in MLLA is also implemented by cutting the 
  development of the parton cascade at different scales $Q_0$, with $Q_0$ 
  chosen equal to the hadron mass. This requires using a slightly more 
  elaborate form of the spectrum than Eq.~(\ref{limit-spectrum}), see 
  Ref.~\cite{Dokshitzer:1988bq}, and leads to an energy-independent stiffening 
  of spectra for heavier hadrons. The comparison to {\em inclusive\/} yields in 
  elementary-particle collisions were done with a unique cutoff parameter for 
  all particle types, $Q_0=\Lambda_{\rm eff}=253$~MeV.}
This resummation and the LPHD prescription give a good description of the 
measured longitudinal distributions of hadrons $D^h(x,Q^2)$ over a wide energy 
range, both in $e^+e^-$~\cite{Braunschweig:1990yd,Abbiendi:2002mj} and in 
$p\bar p$~\cite{Acosta:2002gg} collisions. 
For instance, Fig.~\ref{fig:hump-backed_plateau} shows the distribution of 
inclusive hadrons $D^h(x,Q^2)$ inside 17.5~GeV jets, a typical jet energy at 
RHIC, as measured in $e^+e^-$ annihilation by the TASSO Collaboration~\cite{%
  Braunschweig:1990yd}, together with the MLLA prediction with $K^h=1.35$.

The formalism developed in Ref.~\cite{Borghini:2005em} to describe the 
medium-induced distortion of jets reduces to the MLLA baseline in the absence 
of a medium. 
This new approach involves different approximations, emphasizing other physical 
aspects, from the standard models of parton energy loss that are currently used 
in the phenomenology of RHIC data. 
Thus, present model comparisons to RHIC data start with a medium-modified 
energy spectrum of radiated gluons, 
$\dd I^{\rm tot} = \dd I^{\rm vac} + \dd I^{\rm med}$~\cite{Baier:2000mf,%
  Gyulassy:2003mc,Kovner:2003zj}. 
The part corresponding to the ``normal'' vacuum radiation shows a double 
logarithmic dependence $\dd I^{\rm vac} = \frac{\alpha_s}{\pi^2} 
  \frac{\dd\omega}{\omega} \frac{\dd{\bf k}}{{\bf k}^2}$; its integral over 
${\bf k}$ gives rise to the leading $\ln Q^2$ term in the DGLAP evolution 
equation. 
This contrasts to the ${\bf k}$-integration of $\dd I^{\rm med}$, which is 
infrared- and ultraviolet-safe~\cite{Salgado:2003gb} and leads to a 
nuclear-enhanced ``higher-twist'' contribution, $\propto\hat{q}L/Q^2$, where 
$\hat q$ is the transport coefficient that characterizes the medium, subleading 
in an expansion in $1/Q^2$, but enhanced with respect to other such terms by a 
factor proportional to the geometrical extension $\sim L$ of the target. 
In practice, however, the parton virtuality does not enter the existing 
comparisons to experimental data, where one rather considers the 
${\bf k}$-integrated gluon distribution 
$\omega\,\frac{\dd I^{\rm med}}{\dd\omega}$, neglecting the $Q^2$-dependence.
In addition, existing approximations only include the extra source of gluon 
radiation $\dd I^{\rm med}$ for the leading parton, dropping it for the further 
medium-induced splittings of subleading partons in the shower. 

The obvious way to improve over this state of the art is to replace the double 
differential gluon spectrum $\dd I^{\rm vac}$ by $\dd I^{\rm tot}$ in {\em all\/} 
leading and subleading splitting processes of a medium-modified parton cascade.
This can only be done within a Monte-Carlo approach, which we intend to develop 
in future studies.
The first step in that direction, which is still fully analytical, consists in 
using an additional approximation: 
instead of using the computed ${\bf k}$-integrated medium-induced distribution 
$\omega\,\frac{\dd I^{\rm med}}{\dd\omega}$, we replaced it by a constant 
$f_{\rm med}$. 
One can check that this is not too crude an assumption in the kinematic regime 
tested at RHIC, where it amounts to a similar uncertainty as that arising from 
whether one should use the multiple-soft scattering approach or the single hard 
scattering picture. 
Under this assumption, we have then used the extra medium-induced spectrum 
$\omega\,\frac{\dd I^{\rm med}}{\dd\omega}$ on the same level as 
$\omega\,\frac{\dd I^{\rm vac}}{\dd\omega}$, i.e., as a leading logarithmic 
correction~\cite{Borghini:2005em}.
With this ansatz, our formalism ensures energy-momentum conservation at each 
parton splitting, and treats all leading and subleading parton splittings on the
same footing. 

\section{Phenomenological predictions}
\label{s:results}

As stated above, in our analytical approach we approximate the medium-induced 
contribution to the gluon spectrum $\omega\,\frac{\dd I^{\rm med}}{\dd\omega}$ by 
a constant $f_{\rm med}$ in the kinematically relevant range of $\omega$. 
This actually amounts to considering ``medium-modified parton splitting 
functions'', which differ from the standard ones by enhancing their singular 
parts by a factor $(1 + f_{\rm med})$.\footnote{%
  Such a modification of parton splitting functions was discussed in 
  Ref.~\cite{Guo:2000nz}, where it results from considering nuclear-enhanced 
  twist-four parton matrix elements in studies of deeply inelastic $eA$ 
  scattering.}
This formulation allows us to follow the same line of technical arguments as 
that used for the calculation of jet multiplicity distributions in the absence 
of a medium~\cite{Dokshitzer:1988bq}.

\subsection{Distortion of the hump-backed plateau}
\label{sec:hump-backed_plateau}

We can thus compute the momentum distribution of partons within a 
(gluon-initiated) parton cascade: 
\begin{eqnarray}
x\bar D_g(x, \tau) &\!=\!&
  \displaystyle \frac{4N_c(1+f_{\rm med})\tau}{b\hat B(\hat B+1)}\cr
 & & \displaystyle\times\!
  \int_{-\ii\infty}^{+\ii\infty}\!\frac{\dd\nu}{2\pi\ii}\, x^{-\nu} 
  \Phi(-\hat A+\hat B+1,\hat B+2;-\nu\tau), \nonumber\\[-2mm]
 & & \label{limit-spectrum}
\end{eqnarray}
where $\tau\equiv\ln(Q/\Lambda_{\rm eff})$, $\Phi$ is the confluent hypergeometric
function, 
\begin{eqnarray*}
 & \displaystyle\hat A\equiv \frac{4N_c(1+f_{\rm med})}{b\nu}, \qquad 
\hat B=\frac{\hat a}{b},& \\
 & \displaystyle\hat a=\frac{11+12f_{\rm med}}{3}N_c+\frac{2}{3}\frac{N_f}{N_c^2}, 
\qquad b=\frac{11}{3}N_c-\frac{2}{3}N_f, &
\end{eqnarray*}
and $N_f$ is the number of active flavors.
Quite naturally, by setting $f_{\rm med}=0$ in Eq.~(\ref{limit-spectrum}) one 
recovers the usual ``vacuum'' expression of $x\bar D_g(x, \tau)$. 

To exemplify the effect of the medium-enhanced gluon radiation on the 
hump-backed plateau of particle production, we compare in 
Fig.~\ref{fig:hump-backed_plateau} the spectra Eq.~(\ref{limit-spectrum}) for a 
jet with energy $E_{\rm jet}=17.5$~GeV, shown as a function of $\ln(1/x)$, in the 
cases $f_{\rm med}=0$ (no medium) and $f_{\rm med}=0.8$, where the choice of the 
latter value will be justified in Sec.~\ref{sec:RAA}.\footnote{%
  Both spectra were actually scaled by a factor $K^h=1.35$, so that the vacuum 
  one match the corresponding TASSO data. This goes along the idea that the 
  medium affects the development of the parton shower, while hadronization takes
  place ``in the vacuum'' and thus the LPHD coefficient should be the same in 
  both cases.}
One clearly sees that the effect of the medium is a strong distortion of the 
distribution, with a depletion of the number of particles at large $x$, and 
correspondingly a largely enhanced emission of particles at small $x$: 
due to energy-momentum conservation in the parton cascade, the energy which in 
the vacuum is taken by a single large-$x$ parton is redistributed over many 
small-$x$ partons in the presence of a medium. 

\begin{figure}[t]
  \includegraphics*[width=\linewidth]{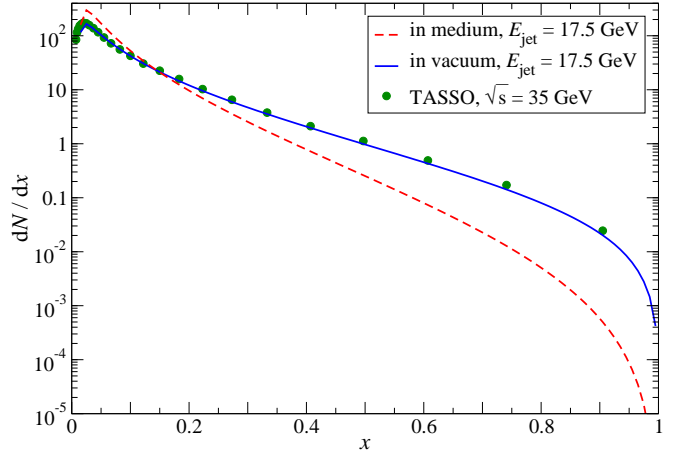}
  \caption{Single inclusive distribution of hadrons, ${\rm d}N/{\rm d}x$, 
    inside a jet of energy $E_{\rm jet}=17.5$~GeV, as a function of 
    $x=E_{\rm jet}$, as measured by TASSO~\cite{Braunschweig:1990yd} and 
    within MLLA (solid curve: $f_{\rm med}=0$; dashed curve: $f_{\rm med}=0.8$).}
  \label{fig:fragmentation-function}
\end{figure}

An alternative way to illustrate the influence of the medium on the spectra is 
to display the distributions as a function of the momentum fraction $x$, see 
Fig.~\ref{fig:fragmentation-function}.
The plot emphasizes the suppression in high-$x$ hadron production, which is the
aspect tested by experimental studies of the high-$p_T$ spectra of hadrons, 
rather than the enhancement in small-$x$ particles.

\subsection{Associated multiplicity}
\label{sec:associated_multiplicity}

\begin{figure}[t]
  \includegraphics*[width=\linewidth]{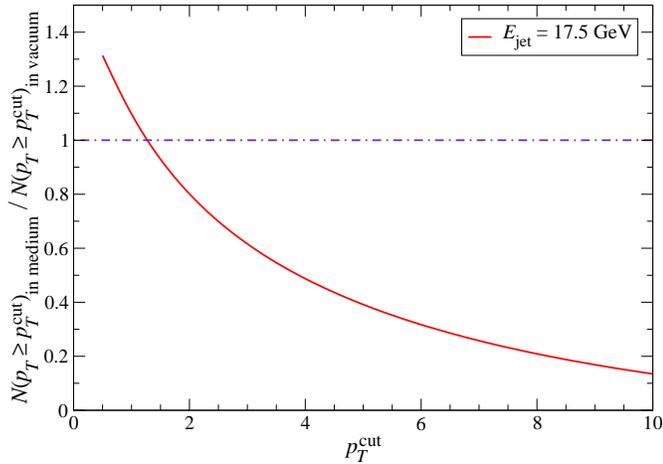}
  \caption{Ratio of the multiplicities of particles above a lower momentum cut 
    $p_T^{\rm cut}$ inside a jets of energy s$E_{\rm jet}=17.5$~GeV developing in a 
    medium ($f_{\rm med}=0.8$) and in the vacuum.}
  \label{fig:associated_multiplicity}
\end{figure}

Once the longitudinal multiplicity distribution inside a jet is known, a 
straightforward integration yields the number of hadrons inside the jet with 
transverse momenta larger than a given cut, ${\cal N}(p_T\geq p_T^{\rm cut})$. 
We can then calculate this multiplicity for jets with the same energy both in 
the presence of medium effects (in which case, the lower cut gives some control 
on the high-multiplicity soft background of nucleus--nucleus collisions at RHIC 
or LHC, over which the jet develops) and in vacuum, and compute their ratio.
We display the latter in Fig.~\ref{fig:associated_multiplicity} for jets with 
$E_{\rm jet}=17.5$~GeV and medium effects modeled by a constant coefficient 
$f_{\rm med}=0.8$.
The ratio is seen to be smaller than 1 for $p_T^{\rm cut}\gtrsim 1.5$~GeV/$c$, 
while the medium-induced enhancement in soft-particle production becomes 
dominant for smaller values of the transverse-momentum cut.
The crossover value is close to that reported by the STAR Collaboration in 
attempts at measuring the excess of particles inside the back jet over the soft 
background~\cite{Adams:2005ph}. 

A more thorough comparison between our calculation and the STAR data would have 
to take into account several effects which we did not consider in our simple 
approach. 
First, in our calculation the energy of the jet is known and fixed, while 
experimentally it spans some range in energy.
In the same line, we did not try to reproduce the experimental fact that the 
high-$p_T$ particles measured at RHIC come from jets that have traversed 
different in-medium path lengths, and thus radiated different energy fractions.
Eventually, we did not model the soft background of Au--Au collisions at RHIC, 
which extends well over the region where the medium-enhanced part of the 
multiplicity lies. 
Nevertheless, the reasonable agreement between our calculation and the 
experimental measurement is a further hint that energy is indeed redistributed 
from high- to low-$x$ partons through the influence of the medium, although 
nothing definitive on the actual shape of the distorted spectrum of radiated 
partons can be claimed at the moment. 
Note, however, that the calculations reported in Ref.~\cite{Borghini:2005em} 
indicate that the crossover between enhancement and depletion should take place
at transverse momenta $p_T^{\rm cut}\sim 4-7$~GeV/$c$ for jets of energy 
$E_{\rm jet}=100-200$~GeV, which should be accessible at the LHC. 
This should leave a window above the upper kinematic boundary of the soft 
background, in which there is an enhancement of the jet multiplicity, thereby 
allowing a more detailed characterization of the medium-enhanced radiation.

\subsection{Nuclear modification factor}
\label{sec:RAA}

As a last phenomenological prediction of our model, let us compute the 
medium-modification factor $R_{AA}$, defined as the ratio of the hadron yield in 
nucleus--nucleus over that in $pp$ collisions, scaled by the number of binary 
collisions. 
For that purpose, we need to supplement a further ingredient to our approach, 
namely the underlying partonic momentum distribution at RHIC energies. 
Since we only want to show that our implementation of radiative parton energy 
loss is able to result in a factor of 5 suppression of the hadron yields, as 
measured experimentally, we shall restrict ourselves to using a power-law parton
spectrum $1/p_T^{\,7}$. 
We do not take into account the kinematic-boundary influence on the shape of 
the $p_T$-spectrum, nor ``initial-state'' effects like shadowing or Cronin 
effect. 
Eventually, as in Sec.~\ref{sec:associated_multiplicity} we do not attempt to 
model the geometry of the medium, nor the fluctuations in the energy radiated 
by partons with a given energy.

\begin{figure}[t]
  \includegraphics*[width=\linewidth]{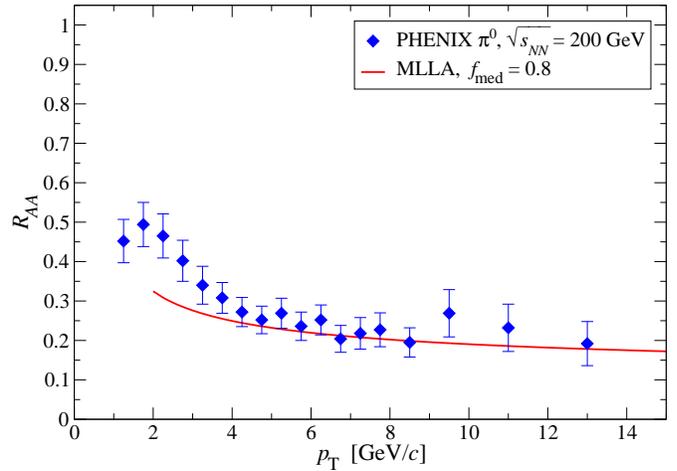}
  \caption{Transverse-momentum dependence of the nuclear modification factor 
    $R_{AA}$, computed for a medium-enhanced radiation of gluons modeled by 
    $f_{\rm med}=0.8$, compared to PHENIX data on $\pi^0$ production~\cite{%
      Adler:2006hu}}
  \label{fig:RAA}
\end{figure}

The result of our calculation is shown in Fig.~\ref{fig:RAA}, together with 
the PHENIX $\pi^0$ data at $\sqrt{s_{NN}}=200$~GeV~\cite{Adler:2006hu}.
We see that our choice of $f_{\rm med}=0.8$ for the coefficient that models 
medium effects allows us to reproduce a modification factor of $\sim 0.2$, which
justifies our using it in all applications reported here. 
This suppression by a factor 5 actually reflects the observation one can make 
on Fig.~\ref{fig:fragmentation-function}, that the spectrum of radiated partons
in a medium is a factor $\sim 3-6$ times smaller than that of partons within a 
jet in vacuum for momentum fractions $x\sim 0.4-0.6$, where $R_{AA}$ mostly 
tests the spectra.
Such values of $x$ are often considered to lie outside the small-$x$ region 
where MLLA is a proper approach to hadron production in elementary-particle
collisions.
Nevertheless, the possible uncertainty of using such a resummation of large 
$\ln(1/x)$ terms (instead of, for instance, fitted fragmentation functions) in 
computing the nuclear modification factor is relatively small, inasmuch as 
$R_{AA}$ is a ratio, which does not depend on the overall normalization of the 
vacuum spectrum, and that we are discussing a suppression by a factor of 5, not 
an effect of a few percent.

\section{Conclusion}

We have reported a first step towards a description of parton cascades 
developing in a medium, which conserves energy-momentum at each successive 
parton splitting, and treats all partons in the shower on the same 
footing~\cite{Borghini:2005em}. 
The simplified analytical formalism we have presented is able to reproduce at 
least semi-quantitatively several characteristic features of RHIC data, such as 
the suppression of high-momentum particle yields and the enhanced soft-particle 
distribution associated to high-$p_T$ trigger particles.
The price we pay for an analytical treatment is the inaccurate handling of the 
$Q^2$-dependence of parton radiation, which is not dealt with more properly in 
other existing formalisms of radiative parton energy loss, and the approximate 
treatment of the $\omega$-dependence of the spectrum, which might not be of 
prime importance in the RHIC regime. 

In the future, we expect to remove the latter deficiencies by implementing our 
approach within a Monte-Carlo formulation. 
In that respect, the analytical results presented here will serve as a powerful 
consistency check for the correct implementation of splitting processes in the 
Monte-Carlo simulation.
Such simulations will allow us to address the issue whether and how 
medium-induced parton energy loss depends on virtuality, as heuristic arguments
suggest it should: since a hard parton of virtuality $Q$ and energy $E$ has a 
lifetime $\sim E/Q^2$ in the rest frame of the dense matter through which it 
propagates, before it branches, high-virtuality partons will branch so quickly 
that their splitting is not influenced by the presence of the medium (so that 
$f_{\rm med}$ should decrease with increasing $Q^2$).
This question will be particularly important for addressing the logarithmically 
wide kinematic range accessible in Pb--Pb collisions at the LHC.
An angular-ordered Monte-Carlo parton cascade will also shed light on other 
issues that have been studied in the absence of medium effects, as two-particle
correlations within jets~\cite{Bassetto:1984ik,Fong:1990ph,Ramos:2006dx}, 
transverse jet broadening, or the differences between light- and heavy-quark
jets~\cite{Dokshitzer:2005ri}. 
Finally, it will also allow one to implement realistic features of the 
collision, as e.g. fluctuations or the geometry of the dense created medium 
traversed by the fast parton. 
\bigskip

{\em Acknowlegements.\/} Fruitful discussions and collaboration with Urs 
Wiedemann are warmly acknowledged.

\end{document}